\documentclass[]{article}

\usepackage{latexsym}

\usepackage{layout}
\usepackage{amsmath}
\usepackage{textcomp}

\title{\bf Geodetic precession and frame dragging observed far from massive objects and close to a gyroscope}

\author{Kostadin Tren\v{c}evski$^1$ and  Emilija G. Celakoska$^2$\\
     $^1$Faculty of Natural Sci. and Math., Ss. Cyril and Methodius Univ., \\
     P.O.Box 162, Skopje, Macedonia, e-mail: kostatre@pmf.ukim.mk\\
     $^2$Faculty of Mechanical Engineering, Ss. Cyril and Methodius Univ.,\\
     P.O.Box 464, Skopje, Macedonia, e-mail: cemil@mf.edu.mk}

\date{}

\begin{document}

\maketitle

\begin{abstract}
Total precession (geodetic precession and frame dragging) depends on the velocity of each source of gravitation, which means that it depends on the choice of the coordinate system. We consider the latter as an anomaly specifically in the Gravity Probe B experiment, 
we investigated it and solved this anomaly. Thus, 
we proved that if our present expression for the geodetic precession is correct, then the frame dragging should be 25\% less than its predicted value.


\noindent Pacs: 04.20.-q; 04.80.Cc; 96.25.De.

\end{abstract}

\pagestyle{myheadings}
\setcounter{equation}{0}
\section{Introduction}\label{sec:1}
\par

The frame dragging problem has been modeled and addressed in  experiments in two ways.  One of the models is described in \cite{Ciu} and it refers to the LAGEOS experiment, while the other is essentially based on Schiff's paper \cite{Sch} and refers to the Gravity Probe B (GPB) experiment. Both experiments describe the behavior of a gyroscope in the vicinity of a rotating source (Earth). In the LAGEOS experiment, the gyroscope is represented by the combination of the nodal longitudes of the two LAGEOS satellites, measuring the orbital shift of the satellites, i.e. the Lense-Thirring effect.  This experiment produced observations corresponding to the prediction of general relativity within 10\%. The GPB experiment measures the shift of the gyroscope's plane directly, and the final measuring results will be discussed in section 5.  
In this paper we consider the model underlying this experiment and we derive that the frame dragging should be 25\% smaller than predicted, given the constraints of the experiment.  Hence the model of the GPB experiment does not correspond to the LAGEOS experiment, and their results and the error estimates cannot be compared.  Specifically, the Lense-Thirring effect is an orbital effect and the LAGEOS experiment tests the equations of motion, while, in the GPB experiment, Fermi-Walker transport is needed to derive the gyroscope's axial precession.  In fact, the GPB experiment tests the applicability of Fermi-Walker transport. In this paper we do not propose any change of the value of the Lense-Thirring effect, but only the value of the frame-dragging. The typical case which allows comparison between the two models (LAGEOS and GPB) is the case of an orbiting gyroscope around a spherical massive rotating body, although both effects (Lense-Thirring and frame dragging) exist more generally. Such an example is the case when the massive body does not rotate, but it moves relative to the chosen coordinate system. In this case the frame dragging alone is not Lorentz covariant but, combined with the geodetic precession, gives a sum that should be Lorentz covariant.

(a) Let us present here notations and some elements of the standard material on this topic.

Let us denote by ${\bf S}$ the spin vector of a gyroscope.  We assume that it is Fermi-Walker transported along the world line.
Differentiating in three dimensions,
\begin{equation}
d{\bf S}/d\tau = {\bf \Omega}\times {\bf S}, \label{1}
\end{equation}
where
\begin{equation}
{\bf \Omega}=-\frac{1}{2}\frac{{\bf v}\times {\bf a}}{c^2} +
\Bigl(\gamma +\frac{1}{2}\Bigr)\frac{{\bf v}\times \nabla U}{c^2} -
\frac{\gamma +1+\frac{\alpha_1}{4}}{4}c\nabla \times {\bf g},
\label{2}
\end{equation}
and ${\bf g}=g_{0i}{\bf e}_i$. As in \cite{AS} we neglect the
symmetric part of the total precession since it is much smaller and periodic for circular orbits. This leaves, as our main concern, the antisymmetric part. Notice that the formula (\ref{2}) is calculated relative to the frame fixed with respect to the distant galaxies \cite{W}.

The first term on the right side in (\ref{2}) denotes the Thomas
precession, which disappears for free-fall orbits. The second term
is the geodetic precession. Notice that the formula for the geodetic
precession was measured to within about 0.36\% using Lunar Laser
Ranging (LLR) data \cite{D,WND,N,WBDF} by considering the Earth-Moon
system as a gyroscope with its axis perpendicular to the orbital
plane. Geodetic precession was recently also confirmed to a
precision of less than 1\% by the GPB experiment.

The third term in (\ref{2}) is the effect of dragging of inertial
frames. We can write $\gamma =1$ and $\alpha_1=0$ according
to general relativity. Then, \cite{W,W1}
\begin{equation}
{\bf \Omega}=2\nabla \times {\bf V},\label{3}
\end{equation}
where
\begin{equation}
V_i=\frac{G}{c^2} \int \frac{\rho 'v'_i}{\mid {\bf x}-{\bf
x}'\mid}d^3x', \quad i=1,2,3.\label{4}
\end{equation}
By solving the corresponding integral,
\begin{equation}
{\bf \Omega}=-\frac{G}{c^2r^3}[{\bf J}-3\hat{{\bf n}}({\bf J} \cdot
\hat{{\bf n}})],\label{5}
\end{equation}
where ${\bf J}$ is the angular momentum of the gravitational
spinning body (the Earth), and $\hat{{\bf n}}$ is the unit radial
vector of the gyroscope. In the case of GPB \cite{BE}, i.e. for
polar motion with constant velocity, this vector averages to
\begin{equation}
\langle {\bf \Omega}\rangle =\frac{G}{2c^2r^3}{\bf J}.\label{6}
\end{equation}

(b) We use the following statements, which are either proven or
acceptable assumptions.

b1. We will use for comparison the known formula ${\bf
\Omega}=(\gamma +\frac{1}{2}) ({\bf v}\times {\nabla U})/c^2$ for
the geodetic precession with respect to the distant stars when the
frame dragging vanishes, as an experimental true.

b2. We assume a weak gravitational field, where the principle of
superposition may be applied.

\pagestyle{myheadings}
\setcounter{equation}{0}
\section{The reason for consideration of the geodetic
precession and the frame dragging}\label{sec:2}

In this paper we will consider only gyroscope in a freefall orbit, so that ${\bf a}=0$. 
As a consequence of (\ref{2}) the angular
velocity is deduced in \cite{W} (eq. (9.5)) in the following form:
$$
{\bf \Omega}= \Bigl(\gamma +\frac{1}{2}\Bigr)\sum_{a}({\bf v}-{\bf
v}_a)\times \nabla \frac{Gm_a}{r_ac^2} -\frac{1}{2}(\gamma
+1)\sum_aG[{\bf J}_a-3\hat{{\bf n}}_a(\hat{{\bf n}}_a\cdot {\bf
J}_a)]/r^3_ac^2$$
\begin{equation}
-\frac{1}{2}\sum_{a}{\bf v}_a\times \nabla \frac{Gm_a}{r_ac^2},
\label{95}
\end{equation}
where ${\bf v}$ is the velocity of the gyroscope, ${\bf v}_a$ is the velocity of the $a$-th spherical body, ${\bf J}_a$ is its angular momentum and $r_a$ is its distance to the gyroscope. Written in this form, the first and the second term in (\ref{95}) are geodetic precession and frame dragging effect respectively and they are Lorentz covariant values, i.e. independent of the choice of coordinate system, while the third term is anomalous since it
depends on the velocity of each body \cite{W}. Thus, the overall expression depends on the choice of the coordinate system 
and hence it is not Lorentz covariant.

The reason why this precession is not Lorentz covariant is the motivation of this paper. We shall not consider many sources of gravitation.   It suffices to consider just one body ($a=1$) with
zero angular momentum, simply as a point mass body, with velocity
$\bf {u}$. This special case leads to the following simple form:
\begin{equation}
{\bf {\Omega}}=\Bigl(\gamma +\frac{1}{2}\Bigr)\frac{{\bf v}\times
{\bf a}}{c^2}-(1+\gamma )\frac{{\bf u}\times {\bf a}}{c^2} =
\Bigl(\gamma +\frac{1}{2}\Bigr) \frac{({\bf v}-\bf {u})\times {\bf
a}}{c^2}- \frac{1}{2}\frac{{\bf u}\times {\bf a}}{c^2}, \label{7}
\end{equation}
where we denote the Newtonian acceleration $\nabla \frac{Gm}{r}$ by
${\bf a}$ in order to simplify the notation.  Notice also that, applying the simple equality (\ref{7}) to each small particle in systems of gravitational bodies while using the principle of superposition for weak gravitational fields (b2), one can deduce (\ref{95}).  So, it is sufficient to study (\ref{7}) in more detail.

The angular velocity is analogous to the magnetic field in electrodynamics. Moreover, it is a part of an antisymmetric tensor which consists of a 3-vector of acceleration and 3-vector of angular velocity, analogously to the tensor of electromagnetic field (\cite{TC,T1,T5,TCbuc,FP}).  We will call this angular velocity the \emph{intrinsic angular velocity}.  The intrinsic angular velocity field is analogous to the field of (Newtonian) acceleration close to the gravitational bodies.  The intrinsic angular velocity includes the {\em proper angular velocity}, a mechanical field, but not resulting from the presence of a moving gravitational body in a close neighborhood.  Our intention is to deal with such coordinate systems for which the proper angular velocity is zero. The following question arises here: How can we recognize these systems; what characteristics they possess? This question will be answered in the next section where we will see that it is possible to give a Lorentz invariant criteria whether a given coordinate system has zero proper angular velocity, i.e. a definition independent of the choice of the observer's coordinate system.

In addition to the intrinsic angular velocity, there is the {\em
exterior angular velocity}. 
It describes how an observer from another coordinate system sees the angular velocity of the chosen coordinate system or the precession of a gyroscope. This is not a field, but it is associated with one object. 

We will see in the next section that, if we measure the precession of a gyroscope by an instrument from a coordinate system with zero proper angular velocity, just close to the gyroscope, then the observed value must be independent of the choice of the coordinate system.  Analogously, the precession of the celestial bodies (i.e. distant stars) in the sky observed from an instrument in a coordinate system with zero proper angular velocity must be independent of the choice of the coordinate system. So now we have two such independent angular velocities, and their difference is also a third such independent angular velocity. This difference is indeed the precession of a gyroscope with respect to the distant stars. 
If the coordinate system of the observer has non-zero proper angular velocity $\psi$, then the first two invariant angular velocities will be changed for the same angular velocity $-\psi$, and their subtraction will remain independent of the choice of the coordinate system.

According to the Mach principle, the axis of a gyroscope remains unchanged with respect to the distant stars.  Presently, this is conceived in such a way that the deviation of the axis with respect to the distant stars is given by (\ref{95}) or (\ref{7}). But the formula (\ref{7}) for this precession will not be satisfactory, because it depends on the choice of coordinate system. Notice that the observation of the precession of
the Earth-Moon axis (de Sitter precession) fits very well with its
prediction because the measurement is done with respect to the
distant stars, and there was no other option than that.
But the measurement of the precession in GPB experiment is much more
sophisticated. Ideally, the gyroscope precession $S$ and
the telescope precession $T$, with the telescope oriented toward the guide star, should be measured separately. However, there are many reasons why
these are not separately measured on long time intervals, and
subtracted afterwards. Instead, in the GPB experiment, the difference $S-T$ is determined every two seconds in a manner that deprives us from having relevant data separately for $S$ and $T$. Although the consideration of the precessions $S$ and $T$ in this paper is analogous to the corresponding precessions in the GPB experiment, they differ from them. Namely, in this paper the proper angular velocity is presumed zero, which is an ideal case of (almost) non-rotating spacecraft. In the GPB experiment, calculations get compounded because the spacecraft rotates.  So the mentioned precessions in the GPB experiment depend on the axis of rotation of the spacecraft and only the difference $S-T$ is relevant, which leads to the precession of the gyroscope with respect to the distant stars.

Since the measurement of the gyroscope precession $S$ gives an inner property of the space-time, it is naturally to be related to the intrinsic metric of space-time. In the next sections we consider the Lorentz invariance of the gyroscope and the telescope precessions in more details.

\pagestyle{myheadings}
\setcounter{equation}{0}
\section{Total precession observed by observers far \\ 
from massive objects}\label{sec:3}

For the rest of the paper, we will mainly focus on exterior angular velocity, which we will simply call angular velocity.

The angular velocity of a gyroscope is not universal, but a relative quantity as the 3-vector of velocity. For example, one test body moves with nonzero velocity vector with respect to one observer (coordinate system), but rests with respect to another observer.  But if we consider the relative velocity of a point $B$ with respect to $A$, then this relative velocity transforms as a 3-vector, neglecting the terms of order $c^{-2}$. We have an analogous situation with the angular velocity.  This value differs between coordinate systems, but the relative angular velocity of one gyroscope with respect to another gyroscope is a quantity which should be (almost) the same according to all observers. We will introduce the notion that one observer sees that another coordinate system rotates with angular velocity $\varphi$ in the following way. If we have two observers from coordinate systems $S_1$ and $S_2$, and assume that the precession of a gyroscope's spin axis is observed to have angular velocity $w_1$ and $w_2$ respectively from $S_1$ and $S_2$, then we accept by definition that the observer from $S_1$ sees that the coordinate system $S_2$ rotates with angular velocity $w_1-w_2$. 
It is important that this definition requires us to assume that the coordinate systems are moving with small velocities with respect to the speed of light and we neglect the terms of order $w/c^2$.  By accepting the previous definition, it is easy to deduce the following conclusions:

1. No observers see any angular velocity of their own coordinate systems.

2. If an observer from $S_1$ sees that $S_2$ rotates with an angular velocity $w$, then the observer from $S_2$ sees that the system $S_1$ rotates with the opposite angular velocity $-w$.

3. If an observer from $S_1$ sees that the coordinate system rotates with an angular velocity $w_1$ and assume that an observer from $S_2$ sees that the coordinate system $S_3$ rotates with an angular velocity $w_2$, then the observer from $S_1$ sees that the coordinate system $S_3$ rotates with the angular velocity $w_1+w_2$. 

The previous discussion shows that although the notion of coordinate system is just a geometric instrument used for calculations, the relative angular velocity between the coordinate systems exists. This conclusion agrees with the Mach principle.

Observing the angular velocity of another coordinate system is affected by the presence of gravitation. For the sake of simplicity, we will at first consider only one gravitational body without angular momentum. It is convenient to consider also observers from inertial coordinate systems far from massive bodies.  Since it is natural to consider that the precession of the coordinate axes observed from the chosen observer does not depend on the velocity of the gyroscope, we introduce the following simple axiom.

{\bf Axiom.} {\em An observer which rests with respect to the gravitational body observes no precession of the coordinate axes of any freely moving coordinate system.} 

Let us now prove the following theorem.

{\bf Theorem.} {\em Let us denote by ${\bf \Omega}_\textrm{gyr.}$ the total spin precession of any gyroscope with respect to observers from inertial coordinate frames far from the massive objects, and let us denote by {\bf u} the velocity of source of gravitation which causes acceleration {\bf a} of the gyroscope. Then:}

(i) ${\bf \Omega}_\textrm{gyr.}-\frac{1}{2}\frac{{\bf v}\times {\bf
a}}{c^2}$ {\em transforms as a spatial 3-vector, neglecting the terms of
order $\vert {\bf \Omega}_\textrm{gyr.}\vert/c^2$, and assuming that the velocities among the observers and the gyroscope are small compared with the light velocity, and}

(ii) {\em the precession of the coordinate axes of freely falling
coordinate system is observed to be} $\frac{1}{2}\frac{{\bf u}\times 
{\bf a}}{c^2}$.

{\bf Proof.}

(i) The theorem is obvious if the coordinates of two observers are
only rotated in the space and they are mutually in rest, because ${\bf
\Omega}_{gyr.}$ and $\frac{1}{2}\frac{{\bf v}\times {\bf a}}{c^2}$
transform as spatial 3-vectors. So, we should consider only two
systems linked by a Lorentz boost, and should prove that the
coordinates of ${\bf \Omega}_{gyr.}-\frac{1}{2}\frac{{\bf v}\times
{\bf a}}{c^2}$ are the same in the two coordinate frames, neglecting
the terms of order $w/c^2$, where $w=\vert{\bf \Omega}_{gyr.}\vert$.  Without loss of generality we can assume only a Lorentz boost along the $x$-axis. In this proof we will assume that the time coordinate is $ict$, for the sake of simplicity.

Assume that the gyroscope moves with a 4-vector of velocity $V_i$.
Let us consider an orthonormal tetrad $A_{i\alpha}$ of 4 vectors,
where $A_{i\alpha}$ is the $i$-th component of the $\alpha$-th
vector. The generality is not lost if we set $A_{i0}=V_i/c$. If the 3-vector of velocity is ${\bf v}=(0,0,0)$, then in the chosen
coordinate frame the corresponding 3-vector $\frac{1}{2}(A'_{32}-A'_{23},A'_{13}-A'_{31},A'_{21}-A'_{12})$ represents angular velocity, where prim denotes differentiation by the time $t$. More generally, if {\bf v} is an arbitrary vector of magnitude $v\ll c$, then the same vector, neglecting the terms of order $\frac{1}{c^2}A'_{ij}$, represents a vector along the axis of rotation in 3 dimensions.

Let us denote by $L_{ij}$ the Lorentz transformation corresponding
to a velocity $v^*$ in the $x$-axis, and let us denote $\bar{A}_{i\alpha}=L_{ij}A_{j\alpha}$. Notice that, neglecting the terms of order $w/c^2$, the derivative with respect to time remains the same in different coordinate frames. Further,
$$\frac{1}{2}(\bar{A}'_{32}-\bar{A}'_{23},\bar{A}'_{13}-\bar{A}'_{31},
\bar{A}'_{21}-\bar{A}'_{12})=$$
$$=\frac{1}{2}(L_{3i}A'_{i2}-L_{2i}A'_{i3},L_{1i}A'_{i3}-L_{3i}A'_{i1},
L_{2i}A'_{i1}-L_{1i}A'_{i2})=$$
$$=\frac{1}{2}(A'_{32}-A'_{23},
L_{11}A'_{13}+L_{10}A'_{03}-A'_{31},A'_{21}-L_{11}A'_{12}-L_{10}A'_{02}).$$

We notice that neglecting the terms of order $w/c^2$ gives
$L_{11}A'_{13}=A'_{13}$ and $L_{11}A'_{12}=A'_{12}$. By
differentiating the equality $A_{ij}A_{i0}=\delta _{j0}=0$ for $j=2$ and $j=3$ we obtain $A'_{ij}A_{i0}+A_{ij}A'_{i0}=0$,
$$A'_{0j}=-(A'_{1j}A_{10}+A'_{2j}A_{20}+A'_{3j}A_{30}
+A_{0j}A'_{00}+A_{1j}A'_{10}+A_{2j}A'_{20}+A_{3j}A'_{30})/A_{00}.$$
Hence, $A'_{02}+i\frac{a_y}{c}\sim wv/c$ and
$A'_{03}+i\frac{a_z}{c}\sim wv/c$. Replacing that $L_{10}\approx
i\frac{v^*}{c}$ and neglecting the terms of order $w/c^2$ we obtain
$$\frac{1}{2}(\bar{A}'_{32}-\bar{A}'_{23},\bar{A}'_{13}-\bar{A}'_{31},
\bar{A}'_{21}-\bar{A}'_{12})=$$
$$=\frac{1}{2}(A'_{32}-A'_{23},A'_{13}-A'_{31},
A'_{21}-A'_{12})+\frac{1}{2}(0,\frac{v^*a_z}{c^2},
-\frac{v^*a_y}{c^2}),$$
$$\bar {\bf \Omega}_{gyr.}={\bf \Omega}_{gyr.}-
\frac{1}{2}\frac{{\bf v}^*\times {\bf a}}{c^2},$$ where ${\bf
v}^*=(v^*,0,0)$. If we denote by {\bf v} and $\bar {\bf v}$ the
velocity vectors of the gyroscope according to the both coordinate
systems and use that ${\bf v}^*={\bf v}-\bar{\bf v}$, we obtain
finally
$$\bar {\bf \Omega}_{gyr.}={\bf \Omega}_{gyr.}-
\frac{1}{2}\frac{({\bf v}-\bar {\bf v})\times {\bf a}}{c^2},$$
$$\bar {\bf \Omega}_{gyr.}-\frac{1}{2}\frac{\bar{\bf v}\times {\bf a}}{c^2}
={\bf \Omega}_{gyr.}-\frac{1}{2}\frac{{\bf v}\times {\bf a}}{c^2},$$
and the proof of (i) is finished.

Since $-\frac{1}{2}\frac{{\bf v}\times {\bf a}}{c^2}$ is the Thomas precession ${\bf \Omega}_\textrm{Thomas}$, part (i) of the theorem states that
$${\bf \Omega}_{gyr.}+{\bf \Omega}_\textrm{Thomas}$$
transforms as a 3-vector. So, the role of the Thomas precession is to complete the precession up to a 3-vector in all systems.

Notice also that in the previous proof we can also replace ${\bf
v}^*={\bf u}-\bar{\bf u}$ instead of ${\bf v}^*={\bf v}-\bar{\bf
v}$, where {\bf u} and $\bar{\bf u}$ are the velocity vectors of the gravitation source with respect to the chosen coordinate systems, and hence we would obtain
$$\bar {\bf \Omega}_\textrm{gyr.}-\frac{1}{2}\frac{\bar{\bf u}\times {\bf a}}{c^2}
={\bf \Omega}_\textrm{gyr.}-\frac{1}{2}\frac{{\bf u}\times {\bf a}}{c^2},$$
which means that ${\bf \Omega}_\textrm{gyr.}-\frac{1}{2}\frac{{\bf u}\times
{\bf a}}{c^2}$ is Lorentz invariant. This is also true because
$\frac{1}{2}\frac{{\bf v}\times {\bf a}}{c^2} -\frac{1}{2}\frac{{\bf
u}\times {\bf a}}{c^2}= \frac{1}{2}\frac{({\bf v}-{\bf u})\times
{\bf a}}{c^2}$ transforms as a spatial 3-vector.
\par

(ii) According to the above axiom, the precession of the coordinate axes observed by the chosen observer does not depend on the velocity of the gyroscope. Thus, we can simply repeat the proof of (i) and replace ${\bf v}^*={\bf u}-\bar{\bf u}$ instead of ${\bf v}^*={\bf v}-\bar{\bf v}$ and for the precession of the coordinate system as a gyroscope we obtain
$$\bar {\bf \Omega}_\textrm{coord.}
-\frac{1}{2}\frac{\bar{\bf u}\times {\bf a}}{c^2} ={\bf
\Omega}_\textrm{coord.}-\frac{1}{2}\frac{{\bf u}\times {\bf a}}{c^2}.$$
Here, the angular velocities $\bar {\bf \Omega}_\textrm{coord.}$ and ${\bf
\Omega}_\textrm{coord.}$ are the observed precession of the coordinate
axes. The 3-vector $\bar {\bf
\Omega}_\textrm{coord.}-\frac{1}{2}\frac{\bar{\bf u}\times {\bf a}}{c^2}$,
i.e. ${\bf \Omega}_\textrm{coord.}-\frac{1}{2}\frac{{\bf u}\times {\bf
a}}{c^2}$ is a zero vector according to above axiom. Hence, we
obtain that the observed angular velocity is ${\bf
\Omega}_\textrm{coord.}=\frac{1}{2}\frac{{\bf u}\times {\bf a}}{c^2}$, and
the proof of (ii) is finished.

\pagestyle{myheadings}
\setcounter{equation}{0}
\section{Total precession observed from an instrument close to the gyroscope}\label{sec:4}

According to Theorem (i), the vector ${\bf
\Omega}-\frac{1}{2}\frac{{\bf v}\times {\bf a}}{c^2}$ transforms as a spatial 3-vector, so assuming that it is a linear combination of $\frac{{\bf u}\times {\bf a}}{c^2}$ and $\frac{{\bf v}\times {\bf a}}{c^2}$ as it is the case in (\ref{7}),
then it must be of the form $k\frac{({\bf v}-{\bf u})\times {\bf
a}}{c^2}$, i.e.
\begin{equation}
{\bf \Omega}_{gyr.}=\frac{1}{2}\frac{{\bf v}\times {\bf a}}{c^2}+
k\frac{({\bf v}-{\bf u})\times {\bf a}}{c^2},\label{9}
\end{equation}
where $k$ is unknown constant. Further we will find that
$k=\frac{3}{2}$ using (b1). 

Notice that, in any inertial coordinate system, the difference between the observed precession of the gyroscope (\ref{9}) and the precession of the coordinate axes $\frac{1}{2}\frac{{\bf u}\times {\bf a}}{c^2}$ must be the same in all inertial coordinate systems, because it is a relative precession of the gyroscope with respect to the coordinate axes. So it must be Lorentz invariant (i.e. transforms as a 3-vector) and it must be the
angular velocity of the gyroscope which is calculated and observed
near the gyroscope. This difference is
$$\frac{1}{2}\frac{{\bf v}\times {\bf a}}{c^2}+
k\frac{({\bf v}-{\bf u})\times {\bf a}}{c^2}- \frac{1}{2}\frac{{\bf
u}\times {\bf a}}{c^2},$$ i.e.
\begin{equation}
\Bigl (k+\frac{1}{2}\Bigr)\frac{({\bf v}-{\bf u})\times {\bf
a}}{c^2}.\label{10}
\end{equation}
So the observed precession of the gyroscopes from the chosen
coordinate system close to the gyroscope and without proper angular
velocity is
\begin{equation}
{\bf \Omega}_\textrm{gyr.}= \sum_i \Bigl (k+\frac{1}{2}\Bigr)\frac{({\bf
v}-{\bf u}_i)\times {\bf a}_i}{c^2}.\label{11}
\end{equation}

Now we should consider the apparent precession (not the true one) of distant stars.  Firstly, let us consider only one gravitational body, as previously.  Then, from an inertial coordinate system which rests with respect to the observed coordinate system, this body rotates with angular velocity
$$\frac{1}{2}\frac{({\bf u}-{\bf v})\times {\bf a}}{c^2},$$
because the relative velocity of the source of gravitation with
respect to the gyroscope is ${\bf u}-{\bf v}$. After the summation
of all particles in the gravitational bodies the observed precession of the coordinate axes is
$$
\sum_i \frac{1}{2}\frac{({\bf u}_i-{\bf v})\times {\bf a}_i}{c^2}.
$$
Hence, observed from a coordinate system without proper angular velocity, all celestial bodies on the sky seem to rotate with the opposite angular velocity, i.e.
\begin{equation}
{\bf \Omega}_{tel.}=\sum_i \frac{1}{2}\frac{({\bf v}-{\bf
u}_i)\times {\bf a}_i}{c^2}. \label{12}
\end{equation}

Earlier, we assumed that all of the coordinate systems have zero
proper angular velocity.  In these coordinate systems, then, all celestial bodies on the sky seem to rotate with the angular velocity (\ref{12}).  We may accept this as a {\em
criterion for one system to have zero proper angular velocity}. We
would like to emphasize that in such rotating coordinate system the rotational accelerations (Coriolis, centripetal, etc.) {\em do not exist}, and also the center of rotation does not exist. It is only a field, like a Newtonian field of acceleration. Perhaps the reason that the Thomas precession disappears in freefall is a consequence of the absence of rotating forces in this case.

Finally, according to (\ref{11}) and (\ref{12}), the precession of the gyroscope with respect to the distant stars is given by
\begin{equation}
{\bf \Omega}_\textrm{rel.}={\bf \Omega}_\textrm{gyr.}-{\bf \Omega}_\textrm{tel.}= \sum_i
k\frac{({\bf v}-{\bf u}_i)\times {\bf a}_i}{c^2},\label{13}
\end{equation}
and hence, according to (b1), it is obvious that $k=\frac{3}{2}$.

We mentioned in section 2 that starting from (\ref{7}), formula
(\ref{95}) can be deduced. Now starting from (\ref{11}), (\ref{12}) and (\ref{13}) we can obtain that the corresponding Lorentz invariant formulas 
for the gyroscope, telescope and the relative precession of the gyroscope with respect to the distant stars are
\begin{equation}
{\bf \Omega}_\textrm{gyr.}=2\sum_{a}({\bf v}-{\bf v}_a)\times \nabla
\frac{Gm_a}{r_ac^2} - \sum_aG[{\bf J}_a-3\hat{{\bf n}}_a(\hat{{\bf
n}}_a\cdot {\bf J}_a)]/r^3_ac^2, \label{11'}
\end{equation}
\begin{equation}
{\bf \Omega}_\textrm{tel.}=\frac{1}{2}\sum_{a}({\bf v}-{\bf v}_a)\times
\nabla \frac{Gm_a}{r_ac^2} -\frac{1}{4} \sum_aG[{\bf J}_a-3\hat{{\bf
n}}_a(\hat{{\bf n}}_a\cdot {\bf J}_a)]/r^3_ac^2, \label{12'}
\end{equation}
\begin{equation}
{\bf \Omega}_\textrm{rel.}=\frac{3}{2}\sum_{a}({\bf v}-{\bf v}_a)\times
\nabla \frac{Gm_a}{r_ac^2} -\frac{3}{4} \sum_aG[{\bf J}_a-3\hat{{\bf
n}}_a(\hat{{\bf n}}_a\cdot {\bf J}_a)]/r^3_ac^2. \label{13'}
\end{equation}

\pagestyle{myheadings}
\setcounter{equation}{0}
\section{Conclusion}\label{sec:5}

At the end we discuss the previous results. Comparing the formula
(\ref{13'}) and the old formula (\ref{95}), we see that the geodetic precession is as expected, but the frame dragging according to (\ref{11'}) is 25\% less than expected according to (\ref{95}).

In the introduction, we explained why the measurements of the LAGEOS experiment do not correspond to the frame dragging in
GPB.  Gyroscope precession can be directly measured in experiments where the gyroscope concretely exists as in the GPB with real gyroscopes, and LLR with the Earth-Moon system as a gyroscope though, in the latter case, the frame dragging caused by the rotation of the Sun is too small to be measured. At present, GPB is the only experiment which can measure the frame dragging. We comment now the final measuring results \cite{GPB-PRL}. 
Among the four gyroscopes in the Gravity Probe B experiment let us 
consider the gyroscope no. 3, for which the geodetic precession is the closest to the predicted value 
-6,606.1 mas/yr, which is without doubt the true value. It is natural to expect that the measurements of the 
frame dragging with this gyroscope are also the most close to the true value. 
The measurements with this gyroscope of the frame dragging 
$(-25.0\pm  12.1)$mas/yr are much closer to the value -29.4 mas/yr predicted by (\ref{13'}), than the 
value -39.2 mas/yr predicted by the General Relativity.  

We emphasize that the calculations in this paper are based on the assumption that the formula for the geodetic precession is correct. If we consider the formula of the geodetic precession, another anomaly appears which is beyond the scope of this paper. Namely, we mentioned that the angular velocity of the gyroscopes is given by (\ref{11'}); it is an inner property which should be derived via the intrinsic metric of the 4-dimensional manifold.
Indeed, all calculations using the intrinsic metric lead to results with respect to the local coordinates, but not with respect to the surrounding space. Obviously, using the metric from general relativity, one can not deduce formula (\ref{11}) for $k=3/2$.  The formula (\ref{11}) for $k=3/2$ is obtained in a forthcoming paper. In that paper, we will consider a parallel transport of an orthonormal frame as whole, but not as separate transport of single vectors such as the velocity vector and the spin vector.  So the mentioned anomalies disappear there. We also note that in a recent paper \cite{TC3} it is shown why the angular velocity $\frac{1}{2}\frac{{\bf u}\times {\bf a}}{c^2}$ (from the Theorem (ii)) leads to Lorentz covariant results for different observers, if we observe orbits.


\begin{thebibliography}{99}

\bibitem{Ciu} I. Ciufolini, E.C. Pavlis,
A confirmation of the General relativistic prediction of the Lense-Thirring effect,
{\em Nature} {\bf 431}, 958--960 (2004).

\bibitem{Sch} L.I. Schiff,
Motion of a gyroscope according to Einstein's theory of gravitation,
{\em Proc. Nat. Acad. of Sci.} USA {\bf 46}(6), 871--882 (1960).

\bibitem{AS} R.J. Adler, A.S. Silbergleit,
General treatment of orbiting gyroscope precession,
{\em Int. J. Theor. Phys.} {\bf 39}(5), 1291--1316 (2000).

\bibitem{W} C.M. Will, Theory and Experiment in
Gravitational Physics (Cambridge University Press, New York, 1993).

\bibitem{D} J.O. Dickey, et al.,
Lunar Laser Ranging: A continuing legacy of the Appolo Program,
{\em Science} {\bf 265}, 482--490 (1994).

\bibitem{WND} J.G. Williams, X.X. Newhall, J.O. Dickey,
Relativity parameters determined from Lunar Laser Ranging,
{\em Phys. Rev. D} {\bf 53}, 6730--6739 (1996).

\bibitem{N} K. Nordtvedt,
Lunar Laser Ranging - A Comprehensive Probe of Post-Newtonian Gravity,
In Proceedings of Villa Mondragone International School of Gravitation and Cosmology, Sept. 2002,
arXiv:gr-qc/0301024.

\bibitem{WBDF} J.G. Williams, D.H. Boggs,  J.O. Dickey, W.M. Folkner,
Lunar laser tests of gravitational physics,
In: R. Ruffini (Ed.), Marcel Grossmann Meeting IX On Recent
Developments in Theoretical and Experimental General Relativity,
Gravitation and Relativistic Field Theories, 2-8 July 2000, Rome,
Italy (World Scientific, 2000) 1797--1798.

\bibitem{W1} C.M. Will,
Covariant calculation of general relativistic effects in an or\-biting gyroscope experiment,
{\em Phys. Rev. D} {\bf 67}, 062003, (2003), arxiv:gr-qc/0212069.

\bibitem{BE} S. Buchman, et al.,
The Gravity Probe B Relativity Mission, 
{\em Advances in Space Res.} {\bf 25}(6), 1177--1180, (2000).

\bibitem{TC} K. Tren\v{c}evski, E.G. Celakoska,
General formulae for frequency shifts using a parallel transport of a 4-wave vector in flat Minkowski space,
World Congress of Engineering, London July 1-3, 2009, Vol.2,
(International Association of Engineers, 2003) 1003--1009.

\bibitem{T1} K. Tren\v{c}evski,
One model of gravitation and mechanics,
{\em Tensor} {\bf 53}, 70--82 (1993).

\bibitem{T5} K. Tren\v{c}evski, V. Balan,
Shrinking of rotational configurations and associated inertial forces,
{\em J. Calcutta Math. Soc.} {\bf 1}(3-4), 165--180 (2005).

\bibitem{TCbuc}K. Tren\v{c}evski, E.G. Celakoska,
On the equations of motion in Minkowski space,
In: C. Udriste (Ed.), International Conference Differential Geometry
- Dynamical Systems - DGDS2007, 5-7 Oct. 2007, Bucharest, Romania,
(Geometry Balkan Press, Bucharest 2008) 211--220.

\bibitem{FP} K. Tren\v{c}evski, E.G. Celakoska, V. Balan,
Research of gravitation in flat Minkowski space,
{\em Int. J. Theor. Phys.} {\bf 50}(1), 1--26 (2011).  

\bibitem{GPB-PRL} C.W.F. Everitt et al., Gravity Probe B: Final Results of a Space Experiment to Test General Relativity,  
{\em Phys. Rev. Lett.} {\bf 106}, 221101 (2011).   

\bibitem{TC3} K. Trencevski, E.G. Celakoska,
Equations of motion for two-body problem according to an observer inside the gravitational field, 
{\em J. Dynamical Systems and Geometrical Theories} {\bf 9}(2), 115--136 (2011). 

\end{thebibliography}
\end{document}